\documentclass[a4paper,fleqn,usenatbib]{mnras}

\usepackage[T1]{fontenc}
\usepackage{ae,aecompl}

\usepackage{graphicx}	
\usepackage{amsmath}	
\usepackage{amssymb}	
\usepackage{tikz}
\usepackage{times}
\usepackage[normalem]{ulem}
\usepackage{multirow}
\usepackage{tabularx}
\usepackage{bm}            
\usepackage{url}
\usepackage{color}

\usepackage{hyperref}

\pdfoutput=1

\newcommand{\FeH}{\ensuremath{[\mathrm{Fe/H}]}}
\newcommand{\dex}{\ensuremath{~\mathrm{dex}}}
\newcommand{\Gyrs}{\ensuremath{~\mathrm{Gyrs}}}
\newcommand{\kpc}{\ensuremath{~\mathrm{kpc}}}

\newcommand{\Msun}{\mathrm{M}_\odot}

\newcommand{\figDir}{fig_pdf/}
\newcommand{\reffig}[1]{Fig. \ref{#1}}

\newcommand{\Eagle}{EAGLE}
\newcommand{\MW}{MW}
\newcommand{\LMC}{LMC}
\newcommand{\SMC}{SMC}
\newcommand{\MWLMC}{MW--LMC}

\usepackage{color}
\definecolor{mycolor}{rgb}{.0,.3,1.}
\newcommand{\changed}[1]{{#1}}

\begin{document}



\title[The MW -- LMC collision]{The aftermath of the Great Collision between our Galaxy and the Large Magellanic Cloud}

\author[]{ Marius~Cautun$^{1}$, Alis~J.~Deason$^{1}$, Carlos~S.~Frenk$^1$ and Stuart McAlpine$^{1,2}$
\\
$^{1}$ Institute of Computational Cosmology, Department of Physics, Durham University, South Road, Durham, DH1 3LE, UK \\
$^{2}$ Department of Physics, University of Helsinki, Gustaf H{\"a}llstr{\"o}min katu 2a, 00560 Helsinki, Finland 
}


\pubyear{2018}

\label{firstpage}
\pagerange{\pageref{firstpage}--\pageref{lastpage}}
\maketitle

\begin{abstract}
  The Milky Way (\MW{}) offers a uniquely detailed view of galactic
  structure and is often regarded as a prototypical spiral galaxy. But
  recent observations indicate that the \MW{} is atypical: it has an
  undersized supermassive black hole at its centre; it is surrounded
  by a very low mass, excessively metal-poor stellar halo; and it has
  an unusually large nearby satellite galaxy, the Large Magellanic
  Cloud (\LMC{}). Here we show that the \LMC{} is on a collision
  course with the \MW{} with which it will merge in
  $2.4^{+1.2}_{-0.8}~\rm{Gyrs}$ (68\% confidence level). This
  catastrophic and long-overdue event will restore the \MW{} to
  normality. Using the \Eagle{} galaxy formation simulation, we show
  that, as a result of the merger, the central supermassive black hole
  will increase in mass by up to a factor of 8. The Galactic stellar
  halo will undergo an equally impressive transformation, becoming 5
  times more massive. The additional stars will come predominantly
  from the disrupted \LMC{}, but a sizeable number will be ejected
  on to the halo from the stellar disc.  The post-merger stellar halo
  will have the median metallicity of the \LMC{}, \FeH{}$=-0.5\dex$,
  which is typical of other galaxies of similar mass to the \MW. At
  the end of this exceptional event, the \MW{} will become a true
  benchmark for spiral galaxies, at least temporarily.
\end{abstract}


\section{Introduction} 
\label{sec:introduction}

The Universe is a dynamical system: galaxies are continuously growing
and undergoing morphological transformation. For the most part, this
is a slow, unremarkable process, but from time to time evolution
accelerates through spectacular galaxy mergers. The Milky Way (\MW{})
appears to have been quiescent for many billions of years but its
demise has been forecast to occur when, in several billion years time,
it collides and fuses with our nearest giant neighbour, the Andromeda
galaxy \citep{van_der_Marel2012}. This generally accepted picture
ignores the enemy within --- the Large Magellanic Cloud (\LMC{}).

The \LMC{} is an unusually bright satellite for a MW-mass galaxy:
observations indicate that only $10\%$ of galaxies of similar mass
have such bright satellites
\citep[e.g.][]{Guo2011,Liu2011,Robotham2012,Wang2012}. While the
\LMC{} has a stellar mass roughly $20$ times smaller than our galaxy
\citep{van_der_Marel2002}, it is thought to possess its own massive
dark halo. Local Group dynamics as well as abundance matching based on
hydrodynamic simulations suggest that the \LMC{} halo mass is around a
quarter of the Galactic halo mass \citep{Penarrubia2016,Shao2018}. A
large total mass is supported indirectly by the complement of satellite
galaxies that the \LMC{} is thought to have brought with it into the
Galaxy. These satellites-of-satellites include the Small Magellanic
Cloud (SMC, the second brightest Galactic dwarf), and a large
fraction of the recently discovered satellites in the Dark Energy Survey
\citep[e.g.][]{Kallivayalil2013,Deason2015,
  Jethwa2016,Sales2017,Kallivayalil2018}.

The atypical brightness of the \LMC{} is just one of several features
that make our galaxy stand out. For its bulge mass, the MW has a
supermassive black hole whose mass is nearly an order of magnitude too
small \citep{Savorgnan2016}. The growth of supermassive black holes
results from a complex interplay between host halo mass, gas supply
and stellar and AGN feedback: in low mass haloes,
${\lesssim}10^{12}\Msun$, stellar feedback is efficient at regulating the
central gas content and black holes hardly grow; in more massive haloes, the stellar feedback driven outflow loses its buoyancy, and stalls, triggering the rapid growth phase of the central black hole
\citep[e.g.][]{Booth2010,Booth2011,Dubois2015,McAlpine2016,Angles-Alcazar2017,Bower2017}. Furthermore,
mergers also play a crucial role by enhancing black hole growth
\citep[e.g.][]{Volonteri2003,Hopkins2005,DiMatteo2008,Goulding2018,McAlpine2018}. Thus,
the low mass of the MW black hole could be a consequence of a low halo
mass and the scarcity of mergers experienced by our galaxy.

The stellar halo of the MW is also atypical, being very metal poor and
of rather low mass
\citep[e.g.][]{Merritt2016,Monachesi2016,Bell2017,Harmsen2017}.
Stellar haloes typically grow through mergers and the tidal disruption
of satellites, and thus provide a unique insight into a galaxy's
assembly history
\citep[e.g.][]{Bullock2005,Cooper2010,Cooper2015,Rodriguez-Gomez2016}.
Dwarf galaxies exhibit a strong correlation between stellar mass and
metallicity \citep[e.g.][]{Kirby2013}, a relation that is reflected in
the stellar haloes of larger galaxies, with higher mass objects being
more metal rich
\citep[e.g.][]{Monachesi2016b,Monachesi2018,DSouza2018a}. Besides the
accreted component, stellar haloes are also predicted to have an
\textit{in situ} component that formed in the main galaxy rather than
in a satellite and was ejected into the halo
\citep[e.g.][]{Brook2004,Zolotov2009,Font2011,Cooper2015,Tissera2013,Pillepich2015}.
However, the significance of this component in the MW is still under
debate \citep[e.g.][]{Helmi2011, Bonaca2017, Deason2017, Haywood2018}.

The low mass and high radial concentration of the Galactic stellar
halo could indicate that the MW has grown slowly through minor mergers
since redshift, $z{\sim}2$, and that its dark matter halo formed early
\citep[e.g.][]{Deason2013,Deason2016,Amorisco2017,Amorisco2017b}. Indeed,
recent studies using \textit{Gaia} astrometric data have shown that
the last major accretion event likely occurred between 8 and $11\Gyrs$
ago, around the time when the Galactic disc was beginning to form 
\citep[][]{Belokurov2018, Helmi2018}.

In this paper we investigate the probable orbital evolution of the
\LMC{} and find that it is on a collision course with the MW. We then
use the state-of-the-art \Eagle{} galaxy formation simulation
\citep{Schaye_15} to predict how the outcome of the \LMC{} merger will
change the MW. In particular, we focus on the evolution of the stellar
halo and the central supermassive black hole of our galaxy, the two
components that make the MW so atypical when compared to other spiral
galaxies of similar stellar mass. Both of these components are known
to be affected by mergers, raising an intriguing question: after the
merger with the \LMC{}, will the MW remain an outlier in so far as its
black hole and stellar halo are concerned? 

Coincidently, our neighbour, Andromeda, presents a very informative
picture of the merger process between a massive dwarf and a MW-sized
galaxy. Andromeda is thought to be in the late stages of such a
merger, in which the Giant Southern Stream and M32 are a tidal stream
and the core of a merging dwarf at least as massive as the LMC
\citep[e.g.][]{Fardal2006,Fardal2013,DSouza2018b}.

This paper is organized as follows: Section~\ref{sec:MW-LMC_future}
presents an orbital evolution model for the Local Group and its
application to the future evolution of the MW--LMC--Andromeda system;
Section~\ref{sec:analogues_Eagle} introduces a sample of \MWLMC{}
analogue systems identified in the \Eagle{} simulation and an analysis
of their evolution; Section~\ref{sec:MW-LMC_merger} offers a
prediction for the post LMC merger properties of the MW central
supermassive black hole and stellar halo; and finally,
Section~\ref{sec:conclusion} presents the conclusions of our study.

\section{The future of the \MWLMC{} system}
\label{sec:MW-LMC_future}
We use a semi-analytic model of the orbital dynamics of the Local
Group to study the future evolution of the \MWLMC{} system. We start
by presenting a detailed description of the orbital model, followed by
the most likely predictions for the evolution of the Local Group.

\subsection{Dynamical model}
\label{subsec:orbital_integration}
We predict the future orbital evolution of the \LMC{} using a
semi-analytic model for the Local Group orbital dynamics, which we
take to be composed of the \MW{}, \LMC{} and Andromeda. The \MW{} and
Andromeda are modelled as having three components: a dark matter halo,
a bulge and a disc, while the \LMC{} is modelled as having only a dark
matter halo and a bulge.  The masses of the various components of the
three galaxies are listed in Table~\ref{table:masses} and correspond
to: MW, LMC and Andromeda halo masses from \citet{Penarrubia2016}; MW bulge
and disc masses from \citet{McMillan2017}; Andromeda bulge and disc
masses from \citet{Savorgnan2016}; and LMC stellar mass from
\citet{van_der_Marel2002}. In particular, our assumed MW halo mass is
in very good agreement with the recent determination using
\textit{Gaia} data by \citet{Callingham2018}, as well as with other
measurements (see e.g. Fig.~7 of
\citeauthor{Callingham2018}). Furthermore, the assumed LMC halo mass
is in good agreement with the estimate by \citet{Shao2018}, as well as
with our own determination based on the \Eagle{} simulation (see
section \ref{subsec:LMC_mass}).

\begin{table}
	\centering
	\caption{ The masses of the dark matter halo ($M_{200}^{\rm
            DM}$), bulge ($M_{\rm bulge}$) and disc ($M_{\rm disc}$)
          of the MW, LMC and Andromeda (M31) used in our orbital model. 
          \changed{We use two dynamical models which differ only by the mass, $M_{200}^{\rm DM}$, assigned to the LMC halo. The \emph{fiducial LMC model}, which corresponds to the \citet{Penarrubia2016} mass determination, is the more realistic one and is the one used for our predictions. The \emph{light LMC model} corresponds to the minimum halo mass given the LMC rotation curve \citep{Gomez2015} and is used to illustrate the effect of a low LMC halo mass. The future evolution of the MW--LMC--Andromeda system for the two models is shown in Figure \ref{fig:orbital_evolution}.}
          The errors are 1$\sigma$ uncertainties and are
          used for calculating the uncertainties in the future
          evolution of the \MWLMC{} system. The halo masses are masses 
          contained within the region whose mean density is 200
          times the critical density. 
        }
 	\label{table:masses}
 	\begin{tabular}{ @{}l ccc@{}} 
 		\hline
 		\hline
 		\\[-0.2cm]
		Galaxy &  $M_{200}^{\rm DM}$  &  $M_{\rm bulge}$  &  $M_{\rm disc}$      \\[.05cm]
  		&  $[~\times 10^{12}\Msun~]$  & $[~\times 10^{10}\Msun~]$ &  $[~\times 10^{10}\Msun~]$    \\ [.10cm]
 		\hline
  		\\[-0.2cm]
 		MW & $1.00^{+0.25}_{-0.25}$ & 1.0 & 4.5  \\[.2cm]
 		M31 & $1.30^{+0.35}_{-0.35}$ & 1.5 & 10.3  \\[.1cm]
 		\hline
        \\[-.1cm]
    	\multicolumn{4}{c}{\bf Fiducial LMC model} \\
 		LMC & $0.25^{+0.09}_{-0.08}$ & 0.27 & --  \\[.1cm]
        \hline
        \\[-.1cm]
    	\multicolumn{4}{c}{\bf \changed{Light LMC model}} \\
 		LMC & $0.05\;\;\;$ & 0.27 & --  \\[.1cm]
        \hline
	\end{tabular}
\end{table}

We model the dark matter halo as a sphere with the Navarro, Frenk \&
White density profile \citep[][hereafter,
NFW]{Navarro1996,Navarro1997}, whose potential is given by:
\begin{equation}
	\Phi_{\rm halo} = -\frac{G M_{200}^{\rm DM}}{r}
    	\frac{ \log\left(1+c \; r/R_{200}\right) }{\log(1+c)-c/(1+c) }
        \label{eq:potential_NFW} \;,
\end{equation}
where $c$ is the concentration parameter, $M_{200}^{\rm DM}$ is the
dark matter halo mass and $R_{200}$ the halo radius. The
concentrations 
of the NFW haloes are taken as the median concentrations for their
mass, which are $c=7$ for the \MW{} and Andromeda, and $c=8$ for the
\LMC{} \citep{Hellwing2016}; the assumed concentration makes little
difference to the model outcome since the uncertainties are dominated
by the halo mass and LMC proper motion errors. The potentials of the
two baryonic components are modelled as a Hernquist bulge
\citep{Hernquist1990},
\begin{equation}
	\Phi_{\rm bulge}=-\frac{GM_{\rm bulge}}{r+r_{\rm c}}
    \label{eq:potential_bulge} \;,
\end{equation}
where $M_{\rm bulge}$ and $r_{\rm c}$ are the bulge mass and scale
radius, respectively,  
and a Miyamoto-Nagai disc \citep{Miyamoto1975}
\begin{equation}
	\Phi_{\rm disk} = -\frac{GM_{\rm disc}}{\sqrt{R^{2}+\left(r_{\rm
      a}+\sqrt{Z^{2}+r_{\rm b}^{2}}\right)^{2}}}
	\label{eq:potential_disc} \;,
\end{equation}
where $M_{\rm disc}$ is the disc mass and $r_{\rm a}$ and $r_{\rm b}$
are the scale lengths. The symbols $R$ and $Z$ denote the radial and
vertical cylindrical coordinates, while $r$ denotes the distance. For
the MW, we take the following constant values of the bulge and disc
scale lengths: $r_{\rm c}=0.7\kpc$, $r_{\rm a}=3.5\kpc$ and $r_{\rm
  b}=0.53\kpc$ \citep{Gomez2015}. For simplicity, we adopt the same
bulge and disc scale lengths for Andromeda; however, these values do
not affect the outcome of the \MWLMC{} merger. 

We implement dynamical friction as a deceleration experienced by the
lower mass galaxy when orbiting within the virial radius of the more
massive companion. We assume that the deceleration is given by
Chandrasekhar's formula \citep{Binney2008},
\begin{equation}
	\frac{{\rm d}{\bf v}}{{\rm d}t} = - \frac{ 4 \pi G^2 M \rho \ln\Lambda } {v^2} 
    	\left[ {\rm erf}(X) - \frac{2X}{\sqrt\pi} e^{-X^2} \right]\frac{{\bf v}}{v}
	\label{eq:chandra_dyn_fric} \;,
\end{equation}
where $M$ is the satellite mass, ${\bf v}$ is the relative velocity of
the satellite and the host halo, $\rho$ denotes the density of the
host at the satellite's position, and $X = v/(\sqrt{2}\sigma)$, with
$\sigma$ the local 1D velocity dispersion of the host halo. We take
the Coulomb factor as $\Lambda=r/\epsilon$, where $r$ is the
instantaneous separation between satellite and host, and $\epsilon$ is
a scale length that depends on the density profile of the
satellite. We take the value of $\epsilon$ from \citet{Jethwa2016} who
performed N-body simulations of the \MWLMC{} systems for a set of MW
and LMC halo mass values. The value of $\epsilon$ that best reproduces
the \LMC{} orbit in the \citeauthor{Jethwa2016} N-body simulations is:
\begin{equation}
	\epsilon =
	\begin{cases}
		2.2 r_s - 14 \kpc & \text{if } r_s \geq 8 \kpc \\
		0.45 r_s     & \text{if } r_s < 8  \kpc
	\end{cases}
    \label{eq:Coulomb_factor_scale_length} \;.
\end{equation}

We position the three galaxies (MW, Andromeda and LMC) at the centre
of their haloes and start the orbit integration using the present day
position and velocities, which we take from the
\citet{McConnachie2012} compilation. When calculating velocities, we
adopt the \citet{Kallivayalil2013} proper motion for the LMC and the
\citet{van_der_Marel2012a} value for Andromeda \citep[note that these
proper motions are consistent with the recent \textit{Gaia} DR2
estimates;][]{Gaia2018,vanderMarel2018}. For each galaxy, we calculate
the gravitational pull exerted by the other two companions and, for
the LMC, we include the additional deceleration due to dynamical
friction (Eqn.~\ref{eq:chandra_dyn_fric}). We then integrate the
equations of motion using a symplectic leapfrog scheme. We define the
\MWLMC{} merger as the moment when the \LMC{} comes within 10\kpc{} of
the \MW{} and, once this has happened, the orbital evolution model
treats the \MWLMC{} system as a single object. This merger threshold
is based on \Eagle{} analogues of the \MWLMC{} system (see
Section~\ref{subsec:selection_MW-LMC}): once the \LMC{}-mass satellite
comes within 10\kpc{} of the central galaxy, it is rapidly tidally
stripped and merges with its central galaxy.

To estimate the uncertainties in the orbit of the \MWLMC{} system, we
Monte Carlo sample the estimates of the \LMC{} proper motions and
distance from the \MW{}, as well as the dark matter halo masses of the
\LMC{}, \MW{} and Andromeda. (See Table~\ref{table:masses} for the
halo masses and associated 1$\sigma$ uncertainties.) We obtain 1000
Monte Carlo realizations of the MW--LMC--Andromeda system and we
calculate the evolution of each realization using the semi-analytic
orbital evolution model.

\subsection{The future evolution of the \MWLMC{} system}
\label{subsec:MW-LMC_future}

\begin{figure}
        \centering
        \includegraphics[width=\linewidth,angle=0]{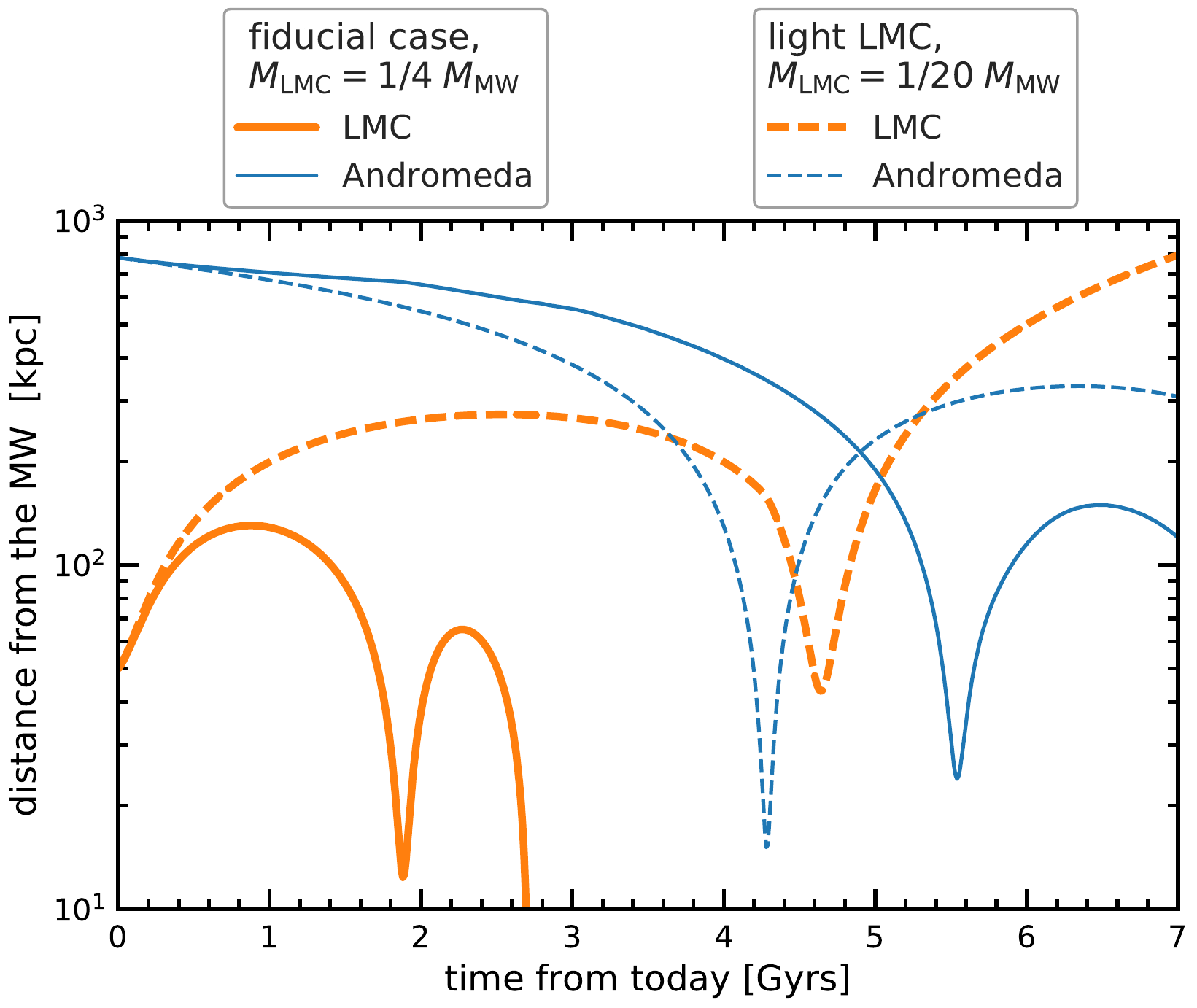}
        \vskip -.2cm
        \caption{ The predicted future evolution of the distance
          between the \LMC{} and the centre of our Galaxy (orange
          lines), and between Andromeda and our Galaxy (blue
          lines). The solid lines correspond to the orbit of our
          fiducial model in which the \LMC{}'s total mass \changed{is given by the \citet{Penarrubia2016} measurement and corresponds to roughly} one
          quarter of the \MW{}'s mass; the dashed lines correspond to the
          orbit for a light \LMC{} halo. The predictions are based on 
          a dynamical model that includes the \MW{}, Andromeda and 
          the \LMC{}, \changed{whose masses are given in Table \ref{table:masses}}. }
          \label{fig:orbital_evolution}
\end{figure}

Fig. \ref{fig:orbital_evolution} shows the future time evolution of
the distance from the MW of the LMC and Andromeda. For our fiducial
model (see Table~\ref{table:masses}), we find that the LMC is on a
radially elongated orbit and will sink towards the Galactic Centre and
merge with the \MW{} in $2.7\Gyrs$. When accounting for observational uncertainties using the Monte Carlos samples described in Section \ref{subsec:orbital_integration}, we find a most likely merger time of 
$2.4^{+1.2}_{-0.8}\Gyrs$ (68\% confidence
level). Furthermore, this merger will take place many
billions of years before the first close encounter between the MW and
Andromeda.  Unlike the forthcoming merger with Andromeda, the
collision with the \LMC{} will not destroy the Galactic disc
\citep[see e.g. the case of the Andromeda--M32 merger discussed
by][]{DSouza2018b} but could still have immense repercussions for the
stellar halo and central supermassive black of our galaxy (see
Section~\ref{sec:MW-LMC_merger}).

Interestingly, if the LMC were much lighter, it would have been on a
very different orbit. To illustrate this,
Fig.~\ref{fig:orbital_evolution} also shows the orbit corresponding to
an LMC halo mass of $5\times10^{10}\Msun$, which corresponds to the
minimum allowed halo mass taking into account the rotation curve of
the LMC and the fact that the LMC extends to at least $15\kpc$
\citep[see e.g.][]{Gomez2015}. A ``light'' LMC would have been on a
harmless, long period orbit and, possibly, could have been kicked out
of the Local Group by the MW--Andromeda merger. However, with a mass
as large as indicated by the recent estimates, dynamical friction due
to the \MW{} mass distribution will cause the LMC to lose energy
leading to a rapid decay of its orbit.

A massive LMC also alters the position and velocity of the barycentre
of the \MWLMC{} system. The change in barycentre affects the orbits of
the other satellites \citep[e.g.][]{Gomez2015,Sohn2017} as well as the
future evolution of the Local Group \citep{Penarrubia2016}. The change
in velocity of the \MWLMC{} barycentre leads to Andromeda having a
larger relative tangential velocity, so that the MW--Andromeda crash
will be less head-on than previously predicted. This will also affect
the time of the first close encounter between the MW and Andromeda and, for example, for our fiducial model show in Fig. \ref{fig:orbital_evolution}, the first encounter will happen in $5.6\Gyrs$, which is $1.3\Gyrs$ latter than predicted for the low LMC halo mass model. The same result holds true even when accounting for observational uncertainties, for which we predict that the first MW--Andromeda encounter will happen in $5.3_{-0.8}^{+0.5}\Gyrs$ (68\% confidence level), nearly ${\sim}1.5\Gyrs$ later than the $3.9^{+0.4}_{-0.3}\Gyrs$ value of
previous estimates \citep{van_der_Marel2012}. Thus, a massive LMC
explains away another puzzle: the apparently anomalously low
transverse velocity of Andromeda, which is very rare in cosmological
simulations \citep{Fattahi2016}. In fact, this is purely a fortuitous
occurrence; the transverse velocity will increase as the \LMC{} moves
along its orbit.

\begin{figure}
        \centering
        \includegraphics[width=1.\linewidth,angle=0]{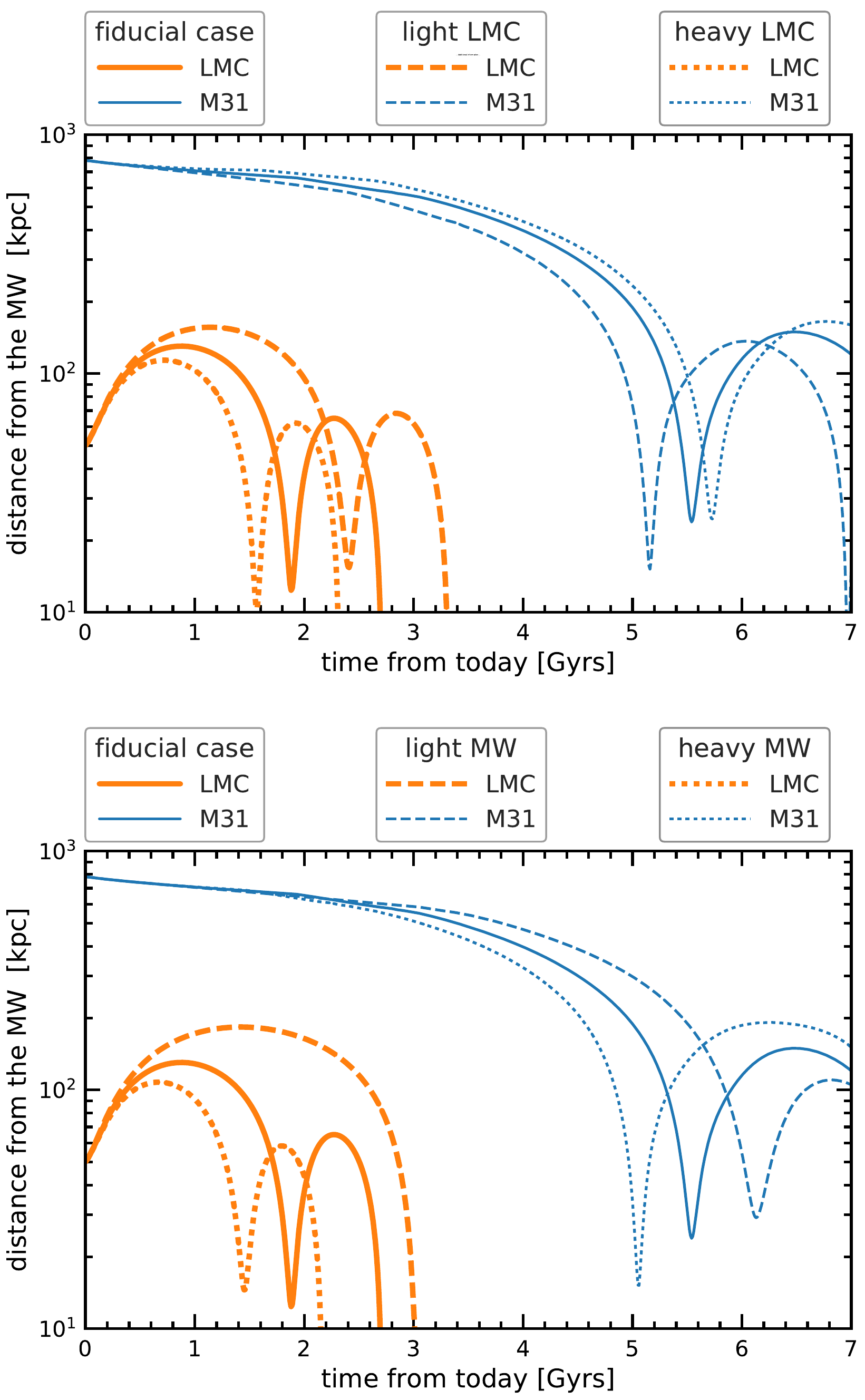}
        \vskip -.2cm
        \caption{The effect of uncertainties in the total halo masses of
          the \LMC{} (top panel) and the \MW{} (bottom panel) on the
          orbits of the \LMC{} and the Andromeda (M31) galaxies. All
          model parameters are kept to their fiducial values (see
          Table \ref{table:masses}) with the exception of the \LMC{}
          (top panel) and \MW{} (bottom panel) halo masses. We show
          the fiducial case as well as results for halo mass values 
          $1\sigma$ above and below the most likely estimate. For
          the \LMC{}, the halo masses are: $2.5$ (fiducial), $1.7$
          (light) and $3.4$ (heavy) $\times10^{11}\Msun$. For the
          \MW{}, the halo masses are:  $1.0$ (fiducial), $0.7$ (light)
          and $1.3$ (heavy) $\times10^{12}\Msun$. 
		}
          \label{fig:orbits_variation}
\end{figure}

Figure~\ref{fig:orbits_variation} illustrates the effect on the
inferred \LMC{} and Andromeda orbits of uncertainties in the halo
masses of the \LMC{} and the \MW{}. A larger \LMC{} or \MW{} halo mass
results in greater dynamical friction and thus a faster merger
timescale for the \MWLMC{} system. Varying the \LMC{} mass within the
$1\sigma$ measurement confidence interval adds an uncertainty of
$\pm0.5\Gyrs$ to the merger time. Varying the \MW{} mass within its
$1\sigma$ range adds a similar level of uncertainty,
$\pm0.4\Gyrs$. Although the \LMC{} and \MW{} halo masses are uncertain
at the ${\sim}30\%$ level, we find that the uncertainty in the merger
time is dominated by measurement errors in the \LMC{} velocity with
respect to the Galactic Centre, which are at the $6\%$ level. When
taking into account all these sources of uncertainty, we predict that
the \MWLMC{} merger will take place in $2.4^{+1.2}_{-0.8}\Gyrs$ (68\%
confidence level).

To sample the measurement uncertainties, we have obtained 1000 Monte
Carlo realizations of the MW--LMC--Andromeda system (see
Section~\ref{subsec:orbital_integration}). These show
that the \MWLMC{} merger is a very likely outcome, with the merger
taking place in $93\%$ of cases. Of the Monte Carlo realizations which
have a \MWLMC{} merger, 90\% have a merger time less than
$4\Gyrs$. Furthermore, in the vast majority of cases (90\%), the
\LMC{} merges with the \MW{} on its second pericentre passage from
today, which corresponds to its third overall pericentre passage as
observations suggest that the \LMC{} is currently just past its first
pericentre \citep{Kallivayalil2013}.

In 92\% of Monte Carlo realizations, the \LMC{} merger takes place
before Andromeda has come within 300\kpc{} of the \MW{}; the presence
of Andromeda at such large distances does not affect the \LMC-\MW{}
merger. This suggests that our omission of M33, which is about twice
the mass of the \LMC{} \citep{McConnachie2012}, from the dynamical
modelling of the Local Group does not affect our conclusions regarding
the \MWLMC{} merger. However, the M33 galaxy does affect the
predictions for the MW--Andromeda collision 
\changed{(for more details see e.g. \citealt{van_der_Marel2012,Patel2017})}. 
Furthermore, there are
additional sources of uncertainty regarding the MW--Andromeda
collision that are not included in our model, such as the uncertainty
in the Andromeda proper motion \citep{van_der_Marel2012}, which is
poorly determined, and the large-scale tidal field in which the Local
Group is embedded (Sawala et al, in prep).

\section{\MWLMC{} analogues in the EAGLE simulation}
\label{sec:analogues_Eagle}
To investigate the outcome of the predicted \MWLMC{} merger, we use
the \Eagle{} suite of cosmological hydrodynamic simulations
\citep{Crain_15,Schaye_15}.  \Eagle{} incorporates the best current
understanding of the physics of galaxy formation, and produces a
realistic population of galaxies with properties that match a plethora
of observations: sizes, star formation rates, gas content and black
hole masses \citep[e.g.][]{Furlong2015,Schaye_15,Trayford2015,Rosas-Guevara2016,Bower2017,McAlpine2017}.

The main \Eagle{} simulation follows the formation and evolution of
galaxies in a periodic cubic volume of $100~\rm{Mpc}$ on a side
assuming the Planck cosmological parameters \citep[][]{planck2014}. It
employs $1504^3$ dark matter particles of mass of
$9.7\times10^{6}M_{\odot}$ and $1504^3$ gas particles of initial mass
of $1.81\times10^{6}M_{\odot}$.  To study analogues of the \MWLMC{}
merger we make use of the \Eagle{} galaxy merger trees built from
${\sim}200$ outputs, which roughly corresponds to one snapshot every
$70~\rm{Myrs}$ \citep{McAlpine2016,Qu2017}.

\subsection{\LMC{} total mass estimates from EAGLE}
\label{subsec:LMC_mass}
According to recent estimates, the dark matter halo of the \LMC{} is
very massive, with a total mass at infall onto the MW of
${\sim}2.5\times10^{11}\Msun$ \citep[e.g.][]{Penarrubia2016}. Here, we
check if this mass measurement is consistent with the \Eagle{}
simulation. The \Eagle{} galaxy mass function at stellar masses of
$10^{9}$ to $10^{10}\ \Msun$ matches observations very well \citep[see
Figure~4 in][]{Schaye_15} and thus \Eagle{} is suitable for inferring
the typical halo mass of LMC-mass galaxies.

To estimate the LMC halo mass at infall, we need to determine the time
when it first crossed the virial radius of the MW and its stellar mass
at that time. We ran the semi-analytic orbit evolution model backwards
to trace the infall orbit of the \LMC{}. Our fiducial model predicts
that the \LMC{} is on first infall \changed{(in agreement with e.g. \citealt{Besla2007,Kallivayalil2013})}, just past its first
pericentre, and that it recently entered the \MW{} halo, having
arrived within $300\kpc{}$ comoving distance of our Galaxy for the
first time only $1.6\Gyrs$ ago. Within the last $2\Gyrs$, the \LMC{}
had an average star formation rate of $0.2\Msun~\mathrm{yr}^{-1}$
\citep{Harris2009}, and thus, at infall $1.6\Gyrs$ ago, the \LMC{} had
a stellar mass of $2.4\times10^9\ \Msun$, which is $10\%$ lower than its
current value.

Under the assumption that, at infall, the \LMC{} was typical of
central galaxies of that stellar mass, we use the redshift, $z=0.13$,
snapshot of the \Eagle{} reference simulation to identify \LMC{}
analogue central galaxies, which we define as having a stellar mass in
the range $2-4\times10^{9}\Msun$. Our selection criterion returned
1714 \LMC{}-mass central galaxies which have a stellar-to-total mass
ratio of $1.23^{+0.53}_{-0.29}\times10^{-2}$ (68\% confidence
level). This corresponds to a LMC halo mass at infall,
$M_{200}=1.9^{+0.7}_{-0.7}\times 10^{11}\Msun$ ($68\%$ confidence
level), about $30\%$ lower, but still consistent to within 1$\sigma$,
with the \citet{Penarrubia2016} measurement.

However, the LMC is not a typical dwarf galaxy. It has a very massive
satellite, the SMC, which is the second largest satellite of the \MW,
with a stellar mass of about a third of the \LMC's
\citep{McConnachie2012}. This motivated us to identify in \Eagle{}
\LMC{}-mass central galaxies that have an \SMC{}-mass satellite. We
define \SMC{}-analogues as satellite galaxies with a stellar mass
between 0.25 and 0.5 times that of their \LMC{}-mass central
galaxy. \LMC{}-mass central galaxies that contain an \SMC{}-mass
satellite are rare; we found only 26 examples in \Eagle. However,
these binary systems are 1.55 times more massive than the typical
\LMC{}-mass central galaxy and have a stellar-to-total mass ratio of
$0.79^{+0.28}_{-0.15}\times10^{-2}$ (68\% confidence level). This
suggests that the \LMC{} is very massive, with a total halo mass of
$M_{200}=3.0^{+0.7}_{-0.8}\times 10^{11}\Msun$ ($68\%$ confidence
level), in good agreement with the \citet{Penarrubia2016} result, but
now roughly $20\%$ higher.

\subsection{Selection of \MWLMC{} analogues}
\label{subsec:selection_MW-LMC}

To identify \MWLMC{} analogues in \Eagle, we started by selecting all
the dark matter haloes with a present-day mass in the range
$(0.5-3.0)\times10^{12}\Msun$, and followed their merger trees to
identify satellite galaxy mergers. We restricted attention to mergers
that took place between 1 and 8 Gyrs ago; the lower bound is needed to
be able to estimate the properties of the system some time after the
merger, while the upper bound corresponds to redshift, $z=1$. We found
that the resulting central black hole mass after the merger was
correlated with several central and satellite galaxy properties. The
black hole grew more when: (1) the merging satellite was more massive;
(2) the central galaxy had more cold gas; (3) the initial black hole
mass was lower; and (4) the merger was not preceded by another
\LMC{}-sized merger within a few gigayears. This and other
considerations motivated us to adopt the following criteria for
identifying analogues that best match the \MWLMC{} system:
\begin{enumerate}
	\item The merging satellite should have a stellar mass in the
          range, $2-4\times10^{9}\Msun$, which corresponds to a small
          interval around the \LMC{} estimated stellar mass of
          $2.7\times10^9\Msun$. 
	\item The central galaxy one dynamical time before the
          merger should have a cold gas mass of at least
          $6\times10^9\ \Msun$, which is motivated by HI and
          molecular gas observations of the \MW{} \citep{Heyer2015}.  
        \item The central galaxy black hole mass one dynamical time
          before the merger should be in the range
          $(2-8)\times10^6\Msun$, which is a factor of 2 either side of
          the value measured for the \MW{} \citep{Boehle2016}.
        \item The merger with the \LMC{} analogue must not have been
          preceded by another merger within the last 5 Gyrs with a
          satellite of stellar mass $1\times10^9\Msun$ or higher. This
          is motivated by the absence of such recent mergers in the
          \MW{}.
\end{enumerate}

The dynamical time provides a characteristic timescale for the merger,
which  increases as the Universe ages and which we take to be the
gravitational free-fall time,
$t_{\textrm{dyn}}=\tfrac{3\pi}{32 G\overline{\rho}}$, where $G$ is the
gravitational constant and $\overline{\rho}$ the mean density of the
system. These selection criteria resulted in 8 \MWLMC{} analogues 
\changed{whose properties are detailed in Table \ref{table:properties_MW-LMC_analogues}. Most analogues correspond to mergers that took place ${\sim}7\Gyrs$ ago, with only one system experiencing a more recent merger at $5\Gyrs$ ago. The early merger time is mainly driven by requiring a close match to the mass of the MW black hole, which is very low when compared to present day galaxies in both observations and the \Eagle{} simulation (see Figures \ref{fig:BH_Mstar_relation} and \ref{fig:time_last_merger}). The MW analogues have a stellar mass a factor of a few lower than our Galaxy; this is because we are studying the progenitors of present day MW-mass galaxies, and, in addition, \Eagle{} underpredicts the central stellar mass of galactic mass haloes by a factor of two \citep[see e.g.][]{Schaye_15}.}

\begin{table*}
	\centering
	\caption{ \changed{Select properties of the eight MW--LMC analogue systems identified in the \Eagle{} simulation. The analogues have the same label as in Figures \ref{fig:evolution_analogues} and \ref{fig:merger_mass_transfer}. The columns give: the lookback time, $t_{\rm merger}$, when the merger took place; the stellar, black hole and cold gas mass, and the bulge-to-total, B/T, ratio of the central galaxy at one dynamical time before the merger; and the maximum stellar mass of the LMC analogue.}
        }
 	\label{table:properties_MW-LMC_analogues}
 	\begin{tabular}{ @{}c cccccc@{}} 
 		\hline
 		\hline
 		\\[-0.2cm]
         & & \multicolumn{4}{c}{MW analogue} & \multicolumn{1}{c}{LMC analogue} \\ [.10cm]
 		\hline
 		\\[-0.2cm]
		Analogue ID & $t_{\rm merger}$ & $M_{\star}$  &  $M_{\rm BH}$  &  $M_{\rm gas}$  &  B/T  &  $M_{\star}$     \\[.05cm]
  		&  $[~{\rm Gyrs}~]$ & $[~\times 10^{10}\Msun~]$  & $[~\times 10^{6}\Msun~]$ &  $[~\times 10^{10}\Msun~]$ &  & $[~\times 10^{9}\Msun~]$   \\ [.10cm]
 		\hline
  		\\[-0.2cm]
 		1 & 7.8 & 1.5 & 4.4 & 1.1 & 0.34 &  3.9  \\
 		2 & 6.1 & 2.0 & 4.2 & 1.0 & 0.29 &  2.8  \\
 		3 & 7.1 & 0.9 & 3.0 & 0.7 & 0.88 &  2.6  \\
 		4 & 7.6 & 1.1 & 5.5 & 0.8 & 0.74 &  4.0  \\
 		5 & 4.8 & 2.6 & 4.3 & 0.9 & 0.36 &  2.5  \\
 		6 & 7.7 & 1.5 & 6.8 & 0.8 & 0.43 &  2.2  \\
 		7 & 7.7 & 1.9 & 4.9 & 1.0 & 0.24 &  3.3  \\
 		8 & 6.1 & 3.3 & 7.4 & 1.0 & 0.13 &  3.8  \\
        \hline
	\end{tabular}
\end{table*}

To disentangle the effects of the merger with an \LMC{} sized
satellite from those due to passive evolution, we selected a control
sample of matched merger-free galaxies. For each \MWLMC{} analogue, we
identified all the galaxies that, in the time interval
$[-2t_{\textrm{dyn}},+2t_{\textrm{dyn}}]$ around the time of the
merger, have not themselves undergone a merger with a satellite of
stellar mass larger than $1\times10^8\Msun$. We further selected only
the top merger-free galaxies that have the closest values of dark
matter halo, stellar, black hole and gas masses to the corresponding
\MW{} analogues. In total, the control sample contains 40 galaxies,
five for each \MWLMC{} analogue system.

\subsection{The merger of \MWLMC{} analogues}
\label{subsec:evolution_eagle}
\begin{figure*}
        \centering
        \includegraphics[width=1.0\linewidth,angle=0]{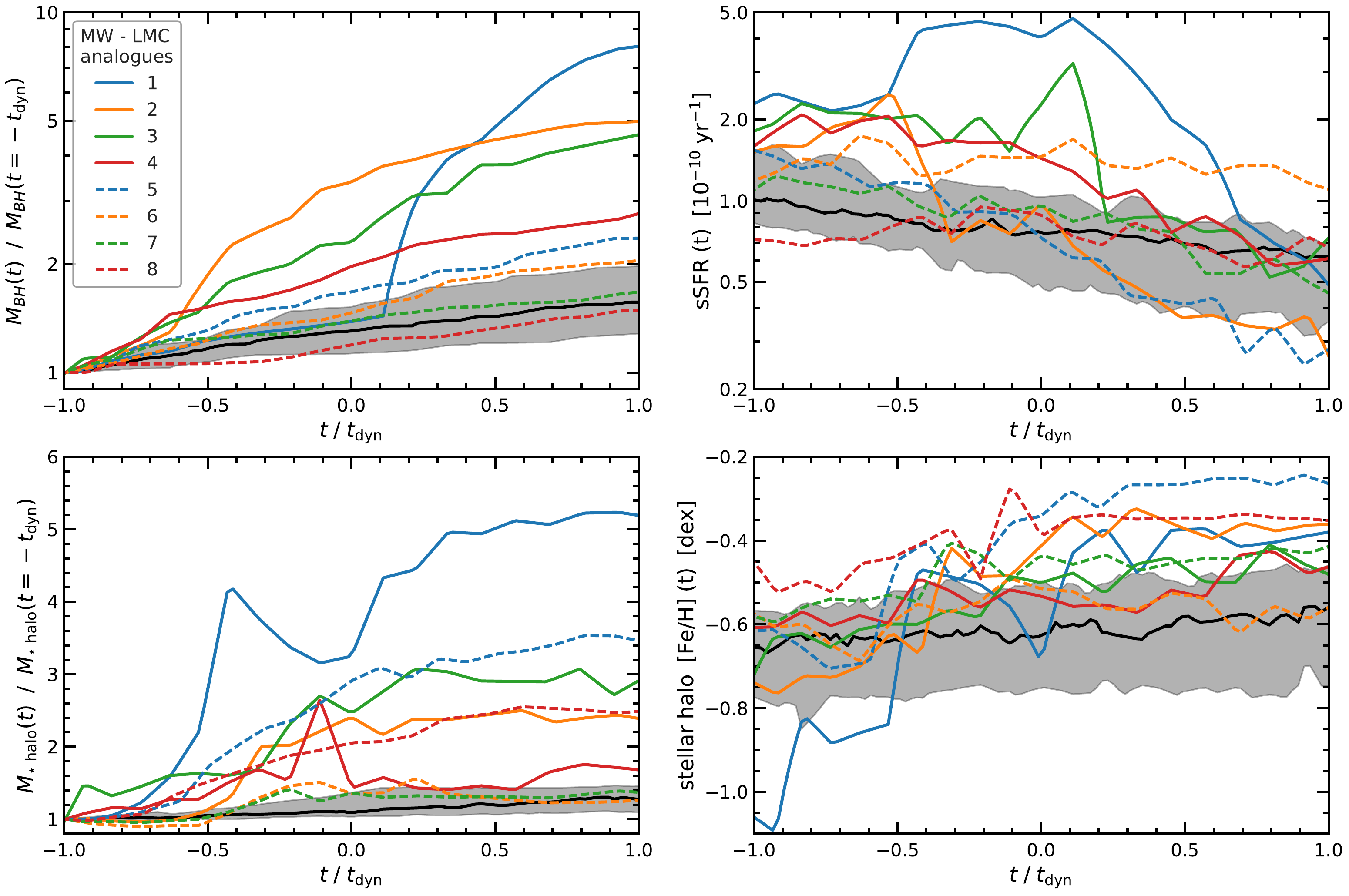}
        \vskip -0.2cm
        \caption{Time evolution around the merger time of the 8
          \MWLMC{} analogues found in the \Eagle{} simulation.  The
          horizontal axis shows time in units of the dynamical time,
          $t_{\textrm{dyn}}$, when the merger took place, with zero
          corresponding to the time of the merger. The panels show:
          the relative increase in central black hole mass (top-left);
          the specific star formation rate of the central galaxy
          (top-right); the relative increase in stellar halo mass
          (bottom-left); and the metallicity of the stellar halo
          (bottom-right). The colour curves show the evolution of the
          8 \MWLMC{} analogues. The black curve and the grey shaded
          region show the median and the 1$\sigma$ scatter for the
          merger-free control sample.  }
        \label{fig:evolution_analogues}
\end{figure*} 

Figure \ref{fig:evolution_analogues} illustrates the evolution of the
8 \MWLMC{} analogues we found in \Eagle{}. Each analogue has a label,
from 1 to 8, with 1 corresponding to the system in which the merger
triggered the largest increase in the mass of the central black hole,
and 8 to the system that experienced the smallest increase in black
hole mass. The top-left panel in Figure \ref{fig:evolution_analogues}
shows the evolution of the central black hole mass from one dynamical
time before the merger to one dynamical time after the merger. The
final black holes ended up having masses between 1.5 to 8 times
(median value 2.6) the initial values. To put this into perspective,
we can compare with the evolution of a control sample of similarly
selected MW-mass galaxies which did not experience any massive
satellite mergers (Section~\ref{subsec:selection_MW-LMC}). Within
the same time frame, the black hole mass of the control sample
increased on average by just a factor of 1.5. This underlines the
critical role of mergers as triggers of black hole growth
\citep[e.g.][]{McAlpine2018}. In particular, this shows that even
minor mergers, in this case with mass ratios around $1:20$, can
trigger significant black hole growth. The enhanced black hole growth
is due to the merger giving rise to asymmetric disturbances in the
central galaxy, which drain angular momentum from the gas of the
central galaxy, and drive it into the centre where the black hole
resides \citep{Mihos1996}. The gas brought in by the merging satellite
represents only a small fraction of the cold gas already present in
the central galaxy and is not the primary driver of the black hole
growth.

The large increase in the central black hole mass of the
\MW{}-analogues raises an important question: would this trigger powerful AGN
activity? 
\changed{In \Eagle{}, the growth of black holes is accompanied by AGN activity, with the injected feedback energy being directly proportional to the recent black hole mass
accretion rate \citep{Schaye_15}. To investigate to what extent AGN activity is enhanced during the \MWLMC{} merger, we calculate the black hole luminosity for both the \MWLMC{}-analogues and the control sample following Eq. 1 from \citet{McAlpine2017}, whereby we assume a radiative efficiency of 10\% \citep{Shakura1973}.}
We find that all 8 analogues show vigorous AGN activity
between one dynamical time before and after the merger. AGNs brighter
than $10^{43}~\mathrm{erg~s}^{-1}$ are active for a fraction of
$0.15-0.40$ of the time, with the highest fraction corresponding to
systems with the largest black hole mass growth. For example, the top
4 \MWLMC{} analogues in terms of black hole mass growth (their black
hole masses increased during the merger by more than a factor of 2.5),
have a $10^{43}~\mathrm{erg~s}^{-1}$ or brighter AGN for a fraction
of $0.3-0.4$ of the time around the merger. 
\changed{This represents a factor of a few times enhancement in AGN activity compared to the control sample, which have similarly bright AGN luminosities for a fraction of only $0.05-0.2$ of the time (see \citealt{McAlpine2018} for a more detailed analysis of merger-induced AGN activity in \Eagle).}

The top-right panel in Figure \ref{fig:evolution_analogues} shows
that, except for one case, the specific star formation rate (sSFR) of
the central galaxy remains roughly flat during the merger. The sSFR,
averaged over the interval of one dynamical time before and after the
merger, takes values from 0.5 to 1.2 (median value 0.75) times the
sSFR at one dynamical time before the merger.  This is in contrast to
the \MW--Andromeda collision, where previous studies predicted that
the star formation rate would roughly double during the merger
\citep{Cox2008}.  Compared to the control sample, which was selected
to have the same amount of cold gas, the \MWLMC{} analogues have
slightly higher star formation rates; this enhancement is seen long
before the actual merger with the \LMC{} analogue. Thus, the
present-day \MW{} could also have a higher sSFR than typical spiral
galaxies, which do not have an \LMC{}-sized satellite. However, we
note that the potential enhancement of the sSFR is relatively weak.

The bottom row in Figure \ref{fig:evolution_analogues} shows the
evolution of the mass and metallicity of the stellar halo during the
merger. To calculate the stellar halo mass, we counted all the star
particles located between 10 and 100$\kpc$ from the central galaxy
\citep{Bell2017} and excluded any stars that were part of bound substructures. The calculation excludes disc stars that are found
beyond $10\kpc$ by removing any star that orbits within $30^\circ$ of
the central disc plane, and then correcting the resulting mass
estimate for the missing angular region by assuming isotropy. The
merger of the \LMC{} satellite analogue can result in a large increase
in stellar halo mass\footnote{The large transient peaks seen in the
  evolution of the stellar halo mass of systems 1 and 4 correspond to
  simulation outputs where the structure finding algorithm wrongly
  assigns most of the satellite mass to the central galaxy, which can
  happen when the satellite is very close to the central galaxy (see \citealt{Qu2017} for more details).}. 
The variation in fractional mass increase is mainly driven by the
variation in the initial mass of the stellar halo, with low mass
systems having the largest fractional mass increase. For example, the
\MWLMC{} analogue labelled number 1 has an initial stellar halo mass
of $5\times10^8\Msun$, approximately equal to the present-day Galactic
stellar halo mass, and its mass is 5 times larger after the
merger.

We also followed the evolution of the stellar halo metallicity at
$30\kpc$ from the central galaxy, which corresponds to the typical
distance at which the metallicity is measured from observations 
\citep[see e.g.][]{Monachesi2016,Bell2017}. To calculate this quantity, we selected
halo stars using the same procedure as for the stellar halo mass
calculation, but now applied to the radial range 20 to 40$\kpc$. The
bottom-right panel in Fig.~\ref{fig:evolution_analogues} shows that an
LMC-mass merger leads to an increase in the stellar halo metallicity,
with the largest increase occuring in the systems with the largest
stellar halo mass growth. The metallicities of dwarf galaxies in
\Eagle{} are too high \citep{Schaye_15} which, in turn, leads to more
metal rich stellar haloes than found in observations (see
Fig.~\ref{fig:stellar_halo_Mstar_relation}). This is not a problem
here since in Fig.~\ref{fig:evolution_analogues} we are only concerned
with relative changes. Furthermore, when extrapolating the \MWLMC{}
analogue results to the real MW, we use the observed metallicities of
the MW and LMC, not the \Eagle{} ones.

\begin{figure}
        \centering
        \includegraphics[width=.9\linewidth,angle=0]{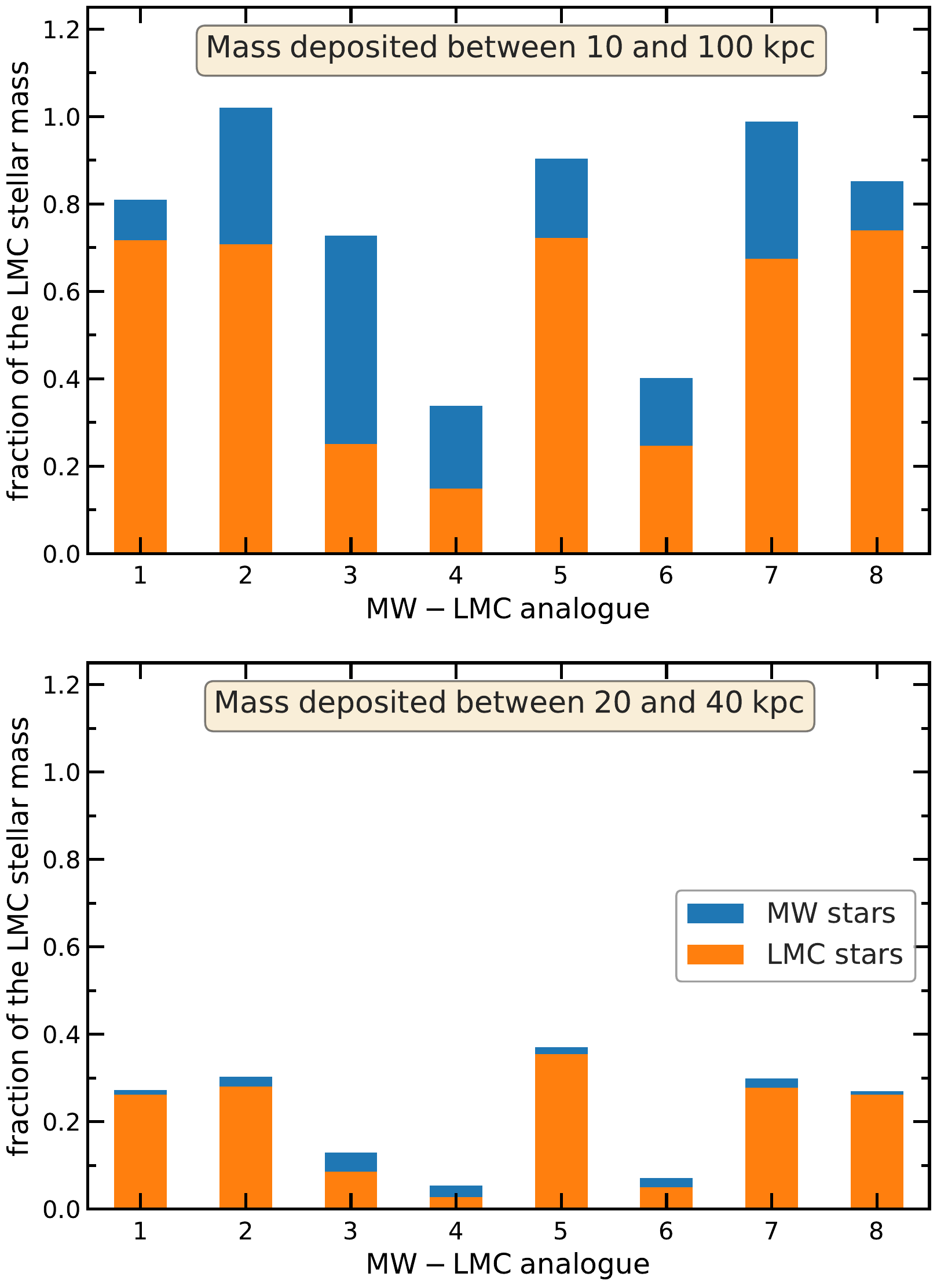}
        \vskip -.2cm
        \caption{The increase in the stellar halo mass expressed as a
          fraction of the stellar mass of the \LMC{} analogue. Plotted
          is the mass deposited in the stellar halo between
          $10-100\kpc{}$ (top panel) and between $20-40\kpc{}$
          (bottom panel). The increase due both to stars that were
          initially part of the \LMC{} analogue (orange) and to stars
          that were initally part of \MW{} analogue (blue) are shown.
          Each column corresponds to a \MWLMC{} analogue, in the same
          order as in \reffig{fig:evolution_analogues}.  }
        \label{fig:merger_mass_transfer}
        \vskip -.2cm
\end{figure}

To make predictions relevant to the actual \MWLMC{} merger, we tracked
the stars belonging to the satellite and central galaxy analogues
(within $10\kpc$ in the latter case), and identified the stars that
one dynamical time after the merger had ended up as part of the
stellar halo. As in the bottom row of
Fig.~\ref{fig:evolution_analogues}, we excluded stars that orbit
within $30^\circ$ of the plane of the central disc, correcting the
resulting mass estimate by assuming that the halo is isotropic. The
results are shown in the top panel of
Fig.~\ref{fig:merger_mass_transfer}, where we express the increase in
the stellar halo mass as a fraction of the \LMC{} analogue stellar
mass. On average, the stellar halo grows by a factor of $0.8$ of the
\LMC{} mass, but the exact values can vary from 0.35 to 1.0.  Most of
the growth results from tidal stripping of the merging satellite, but
there are also central disc stars that are gravitationally ejected
into the halo. In the case of analogue number 3, the mass growth is
dominated by central galaxy stars, but this is more the exception than
the rule.

In the top panel of Fig.~\ref{fig:merger_mass_transfer} three \MWLMC{}
analogue mergers stand out: systems 4 and 6, whose stellar halo grows
only by $0.4$ times the mass of the merging satellite, and system 3,
in which most of the stellar halo mass growth is due to stars kicked
out from the central galaxy. Systems 4 and 6 correspond to mergers in
the plane of the disc and thus, by excluding stars with orbits in the
plane of the disc, we do not take into account the mass deposited within
this region. System 3 corresponds to a bulge dominated central galaxy
(bulge-to-total ratio of $0.88$) and the plane in which the merger
takes place becomes the new plane of the post-merger low-mass
disc. None of these three systems resembles the MW in terms of
bulge-to-disc ratio (${\sim}0.2$ for our Galaxy), or in terms of the
merging satellite orbit (the LMC orbit is nearly perpendicular to the
MW disc). In contrast, the other five \Eagle{} systems are closer
\MWLMC{} analogues: all five central galaxies are disc dominated
(bulge-to-total ratios less than $0.36$) and the growth of the stellar
halo mass in all five is similar to that of \Eagle{} systems 5 and 8,
which have merging satellite orbits nearly perpendicular to the
central disc.

We also calculated the increase in stellar halo mass between $20$ and
$40\kpc$, which is the radial range we used to estimate the increase
in stellar halo metallicity (at $30\kpc$ from the central
galaxy). This is shown in the bottom panel of
Fig.~\ref{fig:merger_mass_transfer}. In contrast to the top panel, the
growth of the stellar halo in this region is much less than in the
inner region, with only $20\%$ of the stellar mass of the \LMC{}
analogue being deposited in the $20-40\kpc$ shell.  Thus, most of the
increase in stellar halo mass takes place in the inner regions
\citep{Cooper2010,Cooper2015,Amorisco2017} and central disc stars are
mainly ejected just outside the $10\kpc$ radius, with very few
reaching a distance of $20\kpc$ or more \citep{Cooper2015}.

\section{The MW before and after the LMC merger}
\label{sec:MW-LMC_merger}
We now investigate the impact of the LMC merger on the mass of the
central black hole and the properties of the stellar halo of our galaxy. For this, we
consider the eight \MWLMC{} analogues identified in the \Eagle{}
simulation that we described in Section~\ref{sec:analogues_Eagle}.

\subsection{The evolution of the central black hole}
\label{subsec:evolution_central_BH}

\begin{figure}
        \centering
        \includegraphics[width=\linewidth,angle=0]{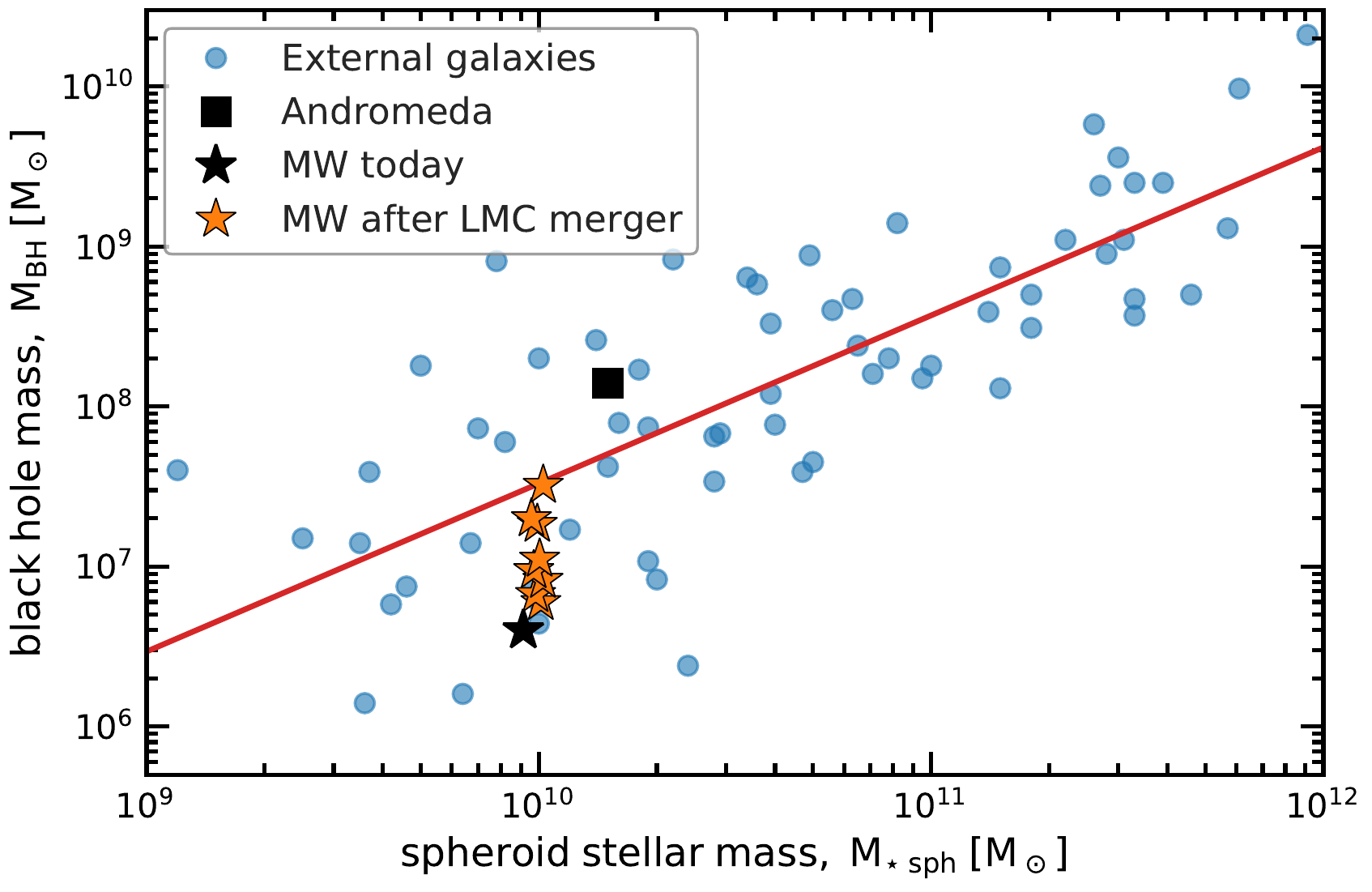}
        \vskip -.2cm
        \caption{The relation between the mass of the central
          supermassive black hole and the stellar mass of the
          spheroid. The relation for a sample of external galaxies
          \citep{Savorgnan2016} is shown by filled circles; the
          current values for the \MW{} and Andromeda are shown by the
          black star and the black square, respectively. The solid
          line is the best-fit power law to the mean relation.
          Measurement errors vary in size from galaxy to galaxy and,
          for clarity, are not shown, but they were used in the
          determination of the mean relation. The orange stars show
          the evolution of the \MWLMC{} analogues identified in the
          \Eagle{} simulation and represent the likely distribution of
          values for the \MW{} after the merger with the \LMC{}.
          }
        \label{fig:BH_Mstar_relation}
\end{figure}

Figure~\ref{fig:BH_Mstar_relation} shows the well-known relationship
between the mass of the central supermassive black hole and the
stellar mass of the spheroid for a large sample of nearby galaxies
\citep[e.g.][]{Kormendy1995,Magorrian1998}.  The large scatter around
the mean trend results from a combination of measurement errors and a
0.5 dex intrinsic scatter \citep{Savorgnan2016}. The black hole of the
\MW{} is plotted as a star. Its mass, $(4.0\pm0.2) \times 10^6 \Msun$
\citep{Boehle2016}, is 8 times smaller than expected from the mean
central black hole - spheroid mass relation. This anomaly is very unlikely to
be due to measurement errors alone: the \MW{} is a $2\sigma$ outlier
in the relation. The lightness of the \MW{} black hole is even more
striking when compared to Andromeda which, for a spheroid that is 1.5
times more massive, has a black hole that is 35 times more massive.

To estimate the MW black hole mass after the LMC merger, we used the
eight \MWLMC{} analogues in the \Eagle{} simulation. In all these
systems, the merger caused a large increase in the mass of the central
black hole, with the post-merger black holes having
masses between 1.5 and 8 times the initial values. We assume that these
mass growth rates are typical of \MWLMC{} mergers, and thus we expect
that the mass of our Galaxy's black hole will increase by a similar 
factor. 

To predict how the Galactic black hole will evolve in
Fig.~\ref{fig:BH_Mstar_relation}, we also estimated the MW spheroid mass post LMC merger in the \Eagle{} analogues.
In the period between one dynamical time before and after the merger,
the average sSFR of the eight \MW{} analogues was 0.5 to 1.2 times the
sSFR at one dynamical time before the merger.  We use these values,
together with the present day sSFR of the MW
\citep{Bland-Hawthorn2016}, $0.03~\textrm{Gyrs}^{-1}$, to estimate the
likely MW stellar mass growth from the present day until the \LMC{}
merger occurs.  On average, in the next $3\Gyrs$, the MW stellar mass
will grow by $7\%$. We also find that the bulge to disc ratio for the eight \MW{} analogues identified in \Eagle{} remains constant during the \LMC{} merger. Thus, the \LMC{} merger will preserve the MW disc and will not lead to a considerable growth of the MW bulge (\citealt{DSouza2018b} have found a similar result for the Andromeda merger with M32). A constant bulge to disc ratio means that the stellar mass growth during the merger is
proportionally split between the two components, and thus the
mass of the bulge and the disc grows by the same factor.

The predicted position in the black hole--spheroid mass diagram of the
post-merger MW is shown in Figure~\ref{fig:BH_Mstar_relation} with
orange star symbols, where each point
\changed{has been scaled according to the growth seen in each of the eight \MWLMC{} analogues identified in \Eagle.}
We find that mergers give rise to
significant black hole growth without a corresponding increase in
spheroid mass. This is exactly the trend needed to bring the \MW{}
black hole into closer agreement with the average black hole--spheroid
mass relation.  Curiously, Andromeda appears to have had a recent,
possibly still ongoing, merger with the massive satellite, M32 
\citep{Fardal2013,DSouza2018b}, that may explain why its black hole is
so much more massive than the \MW's.  In particular, the satellite
merger in Andromeda seems to have left the stellar disc mostly
unharmed, although slightly puffed up, in good agreement with our
finding that the \MWLMC{} merger will not destroy our galactic disc
\citep[e.g. see also][]{Gomez2016,Gomez2017}.

\subsection{The evolution of the stellar halo}
\label{subsec:evolution_stellar_halo}

\begin{figure}
        \centering
        \includegraphics[width=1.0\linewidth,angle=0]{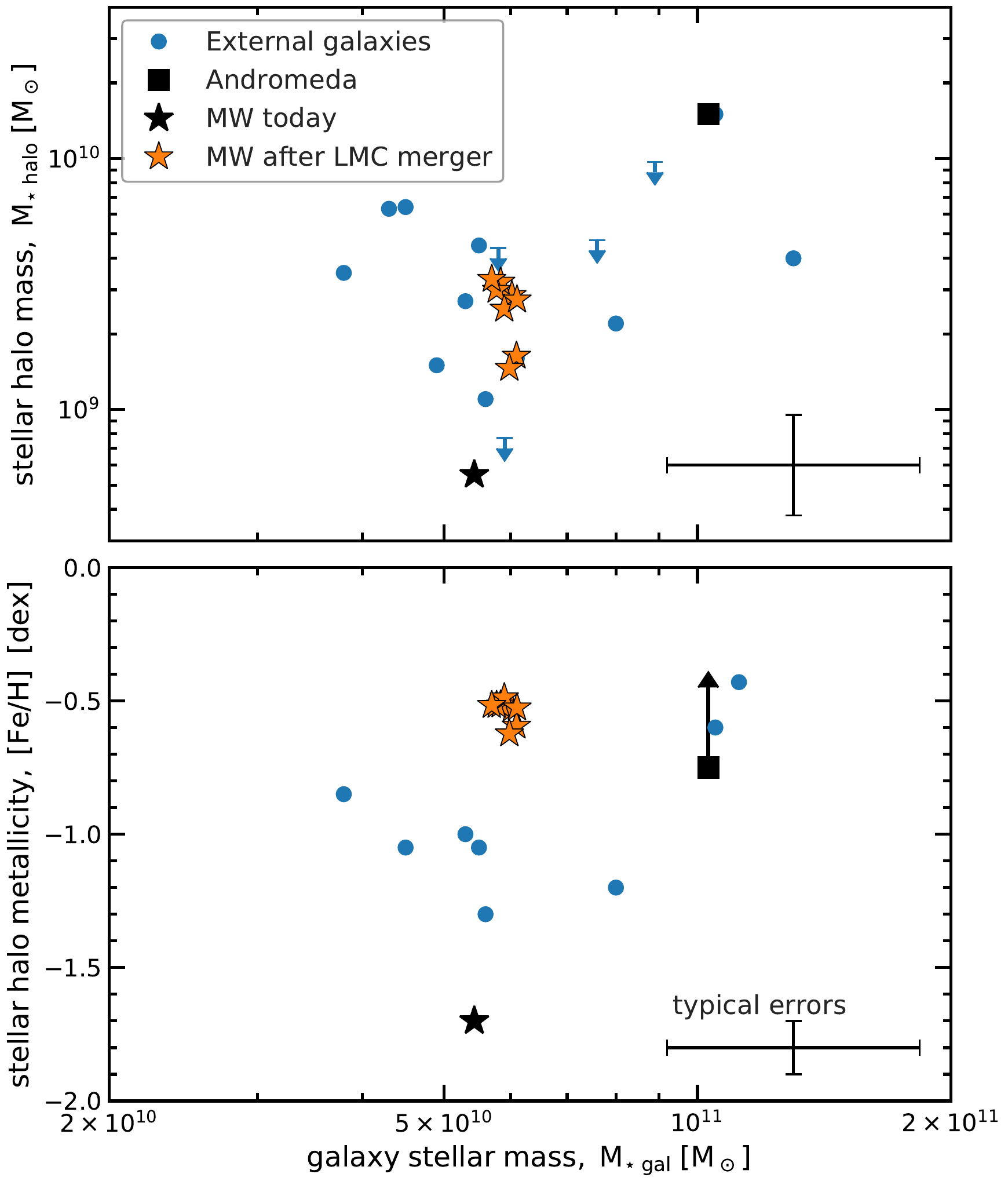}
        \vskip -0.2cm
        \caption{ The relation between the galaxy stellar mass and the
          mass (top panel) and metallicity (bottom panel) of the
          stellar halo for a sample of nearby galaxies
          (filled circles) \citep{Bell2017}. The measured values for the \MW{} and
          Andromeda are shown by a black star and black square,
          respectively. Upper limits on stellar halo mass
           are shown by a downward pointing arrow. The
          square in the right panel shows the metallicity of
          Andromeda's smooth stellar halo; when accounting for
          substructures the median \FeH{} value \citep{Gilbert2014} is
          possibly as high as $-0.4\dex$, as indicated by an
          upward pointing arrow. The orange stars show the likely
          distribution of final locations for the \MW{} after the
          merger with the \LMC{} and are derived from the outcome
          of similar mergers in the \Eagle{} simulation. 
          }
        \label{fig:stellar_halo_Mstar_relation}
\end{figure}

The anomalously low mass and iron abundance of the stellar halo of 
the \MW{} are clearly apparent in
Fig.~\ref{fig:stellar_halo_Mstar_relation} where the properties of
the \MW{} are compared with those of a sample of nearby
galaxies \citep{Bell2017}. There is considerable system-to-system
variation but, most strikingly, the \MW{} is an extreme outlier, with
a very low mass and very metal-poor stellar halo. This is in stark contrast
with Andromeda, which has a particularly massive and metal rich
stellar halo. 

We use the evolution of the stellar haloes of the \MWLMC{} analogues
to predict the expected mass and metallicity of the Galactic halo
after the LMC merger. According to the top panel of
Fig.~\ref{fig:merger_mass_transfer} the merger caused an increase in
the mass of the stellar halo by a factor between 0.35 to 1.0 of the stellar mass of the merging satellite. The mass of
the \MW{}'s stellar halo, $0.55\times10^9\Msun$ \citep{Bell2017}, is
much smaller than the stellar mass of the \LMC{}, $2.7\times10^9\Msun$, so the \LMC{} merger
will result in the Galactic stellar halo becoming 3 to 6 times (median
value 5) times more massive than before the merger. This increase
would place the \MW{} right in the middle of the stellar halo mass
distribution (see top panel of
Fig.~\ref{fig:stellar_halo_Mstar_relation}), turning our galaxy into a
``typical'' object.

\begin{figure}
        \centering
        \includegraphics[width=\linewidth,angle=0]{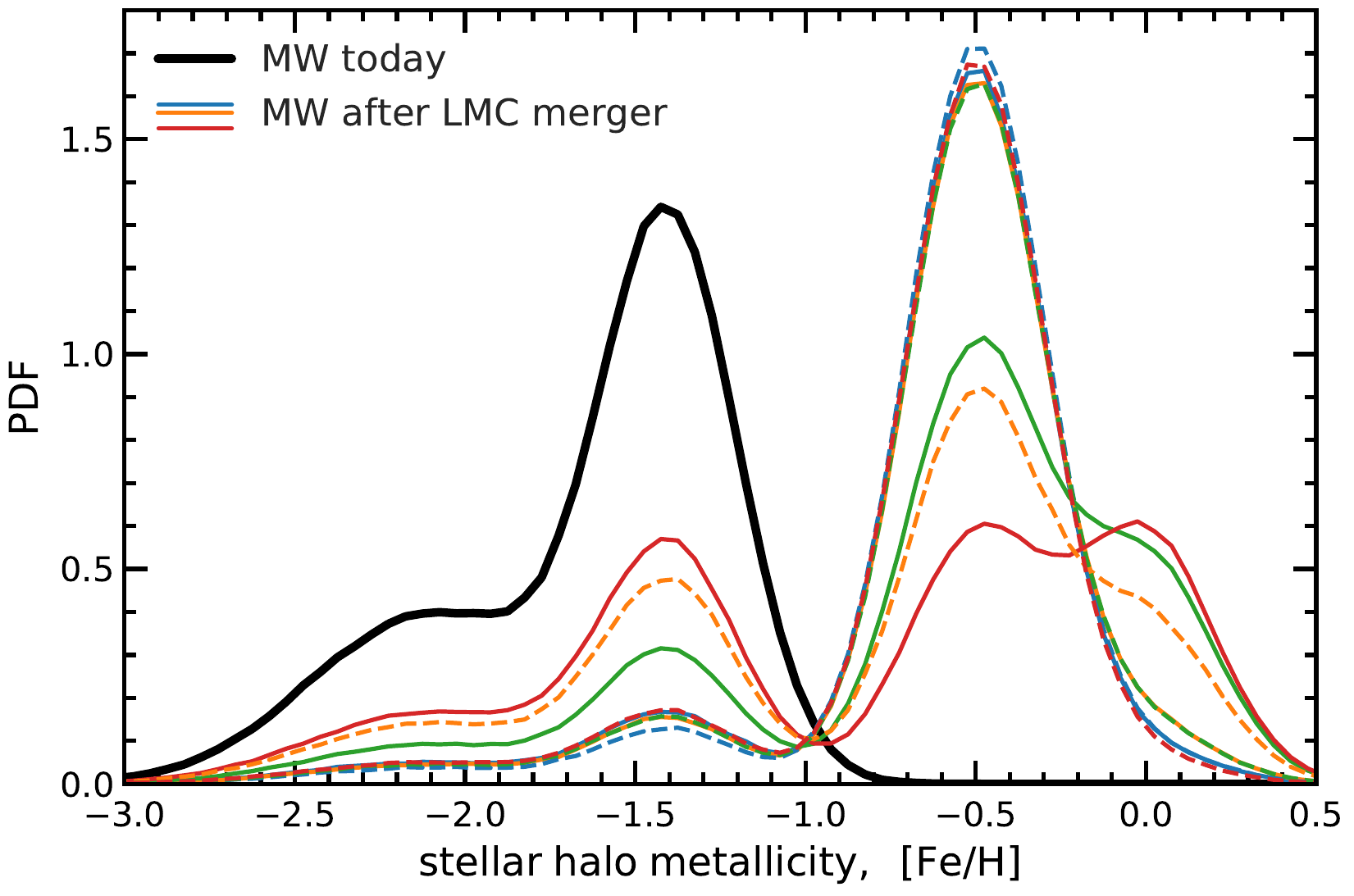}
        \vskip -.2cm
        \caption{The probability distribution function of the
          metallicity of the Galactic stellar halo following the
          merger with the \LMC{}. The present day metallicity
          distribution \citep{Xue2015} is show as a thick black line
          and the possible outcomes after the \LMC{} merger as colour
          lines corresponding to each of the 8 \MWLMC{} analogues; the
          colour-scheme is as in Fig.~\ref{fig:evolution_analogues}. }
        \label{fig:metallicity_distribution}
\end{figure}

\begin{figure*}
        \centering
        \mbox{\hskip -.5cm 
        \includegraphics[width=1.05\linewidth,angle=0]{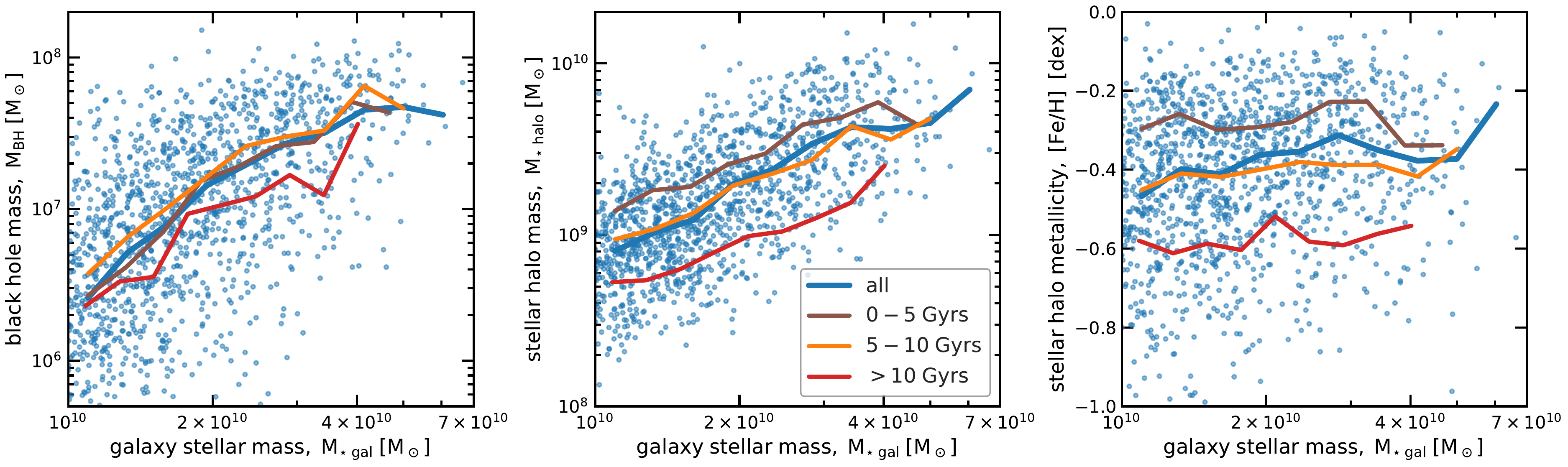}
        }
        \vskip -.1cm
        \caption{The correlation between present-day central galaxy
          properties and the lookback time to the last significant
          merger in the \Eagle{} simulation. The three panels show,
          respectively, the black hole mass, the stellar halo mass and
          the halo metallicity, as a function of the galaxy stellar
          mass. The points show individual \Eagle{} galaxies residing
          in \MW{}-sized dark matter haloes. The lines show the median
          trend for all galaxies (blue) and for subsamples split
          according to the lookback time to the last merger with a
          satellite more massive than the \SMC{}: $0-5\Gyrs$ (brown),
          $5-10\Gyrs$ (orange) and $>10\Gyrs$ (red).  }
        \label{fig:time_last_merger}
\end{figure*}

To predict the metal abundance, \FeH{}, of the post-merger Galactic
stellar halo we use the stellar halo mass growth rates of the \MWLMC{}
analogues at $30\kpc$. We model the post-merger stellar halo as having
three distinct stellar populations: the present-day population of halo
stars, stars stripped from the \LMC{} and stars ejected from the \MW{}
disc. These are mixed according to the mass contributed by each of the
three components. The iron abundance of the present-day halo stars is
well described by two Gaussians with peaks at $-1.4$ and $-2.1$
and widths of $0.2$ and $0.35$, respectively \citep{Xue2015}. This
is shown as the thick solid line in
Fig.~\ref{fig:metallicity_distribution}. The \LMC{} and \MW{} iron
abundances were modelled as Gaussians with $0.2$ dispersion and
mean values of $-0.5$ and $0.0$, respectively
\citep{McConnachie2012,Hayden2015}.

The possible \FeH{} distributions of the \MW{} stellar halo after the
\LMC{} merger are shown in Fig.~\ref{fig:metallicity_distribution},
where each curve corresponds to the weighted sum of the metallicities
of \LMC{} and \MW{} stars at $30\kpc$ inferred from the eight \Eagle{}
\MWLMC{} analogues (see bottom panel of
Fig.~\ref{fig:merger_mass_transfer}). Following the merger, the \LMC{}
stars will dominate the halo and thus the median \FeH{} value will be
close to that of the present-day \LMC{}. However, the distributions
vary somewhat amongst the analogues, with three cases showing a bump
at \FeH{}$=0.0\dex$ corresponding to \MW{} disc stars and also a
sizeable fraction of present-day halo stars. These three cases are the
ones in the bottom panel of Fig.~\ref{fig:merger_mass_transfer}, that
have the smallest increase in stellar halo mass.

The predicted median metallicity of the post-merger Galactic halo is
shown as the set of orange star symbols in the bottom panel of
Fig.~\ref{fig:stellar_halo_Mstar_relation}, each corresponding to one
of the eight \MWLMC{} \Eagle{} analogues.  Since most of the stellar
halo mass growth is due to stripped \LMC{} stars, the median stellar
halo iron abundance is similar to that of the \LMC{},
\FeH{}$=-0.5\dex$, but the exact value varies somewhat from one
\MWLMC{} analogue to another.  The predicted post-merger \MW{} stellar
halo is somewhat more metal rich than the comparison sample of local galaxies. 
\changed{The stellar halo reflects the metallicity of its most massive progenitor; the LMC has had longer to evolve and thus to increase its metallicity than the most massive stellar halo progenitors of other nearby galaxies. Comparing to Andromeda, which experienced a very recent merger, we find that when stellar substructures are included}
the median metallicity of the Andromeda stellar
halo at $30\kpc$ could be as high as $-0.4\dex$ \citep{Gilbert2014}, very similar to our prediction for the post-merger MW.

\subsection{The effect of mergers on \Eagle{} galaxies}
\label{subsec:effect_mergers}

We have shown that even ``minor'' mergers with \LMC{}-sized satellites
can have a large impact on the growth of the central black hole and
the stellar halo of \MW{}-sized galaxies.
In this paper we have argued that the reason why the \MW{} is an
extreme outlier in the properties of these components is a lack of
such satellite mergers in the past.  The growth of a disc as massive
as the \MW{}'s, which contains more than $80\%$ of the total galactic stellar
mass \citep{McMillan2017}, requires a quiet evolutionary history for the
last ${\sim}10\Gyrs$, with no major mergers and, at most, a few minor
ones \citep{Brooks2016r}. 
We can resort to the \Eagle{} simulation to search for correlations
between satellite mergers and the final properties of the central
galaxy.

Figure \ref{fig:time_last_merger} shows the present day distribution
of black hole mass, stellar halo mass and metallicity for \Eagle{}
central galaxies resident in \MW{}-sized dark matter haloes. We split
the sample according to the lookback time to the last merger with a
satellite at least as massive as the SMC (stellar mass
$5\times10^8\Msun$; \citealt{McConnachie2012}).

There is a clear correlation between galaxy properties and lookback
time to the last \SMC{}-like merger. The trend is strongest for the
mass and metallicity of the stellar halo. Fewer recent mergers implies
a lower total number of mergers, a lower stellar halo mass, and a
tendency for the halo to have been built by mergers with low-mass
dwarf galaxies \citep{Deason2016}, which are more metal poor 
than more massive dwarfs \citep{Kirby2013}. The dependency of black hole
mass on the time since the last merger is more complex because the
black hole growth depends not only on the number of mergers, but also
on the amount of gas available in the central galaxy. A more recent
merger means, on average, fewer such mergers at high redshift when
there was more gas available, while no mergers within the last $10\Gyrs$ means fewer
mergers overall, and hence fewer opportunities to trigger black hole
growth.

Fig.~\ref{fig:time_last_merger} shows that a lack of \SMC{}-like
mergers in the past $10\Gyrs$ results in systematically lower black
hole masses and less massive and more metal-poor stellar haloes. This
is consistent with the current state of the \MW{} and suggests that
the lack of significant satellite mergers in the past $10\Gyrs$ is a
likely explanation for why our galaxy is an outlier in scaling
relations \citep[see also][]{Amorisco2017b}. This is consistent with
recent analyses of \textit{Gaia} DR2 data which suggest that today's
\MW{} halo is dominated by stars from a single SMC-mass galaxy which
merged with our galaxy roughly $10\Gyrs$ ago
\citep{Belokurov2018,Helmi2018}.

\section{Discussion and conclusion}
\label{sec:conclusion}

This study was motivated by a desire to understand the physical reason
why the \MW{} is an outlier amongst galaxies of similar stellar mass
in three important properties: the mass of its central black hole, and
the mass and metallicity of its stellar halo. 
\changed{These atypical properties could be due to a lack of significant mergers within the last ${\sim}10\Gyrs$, with observations suggesting that since $z{\sim}2$ the MW stellar halo has grown slowly through minor mergers \citep[e.g.][]{Deason2013,Deason2016,Amorisco2017,Amorisco2017b}. This hypothesis is supported by our analysis of the \Eagle{} hydrodynamical simulation, which predicts that MW-mass galaxies that have not recently experienced a merger with a
galaxy more massive than the SMC have systematically smaller black holes and lower mass and more metal-poor stellar halos, just like our own MW galaxy. 
Furthermore, \citet{Amorisco2017b} showed that hosts that have recently accreted massive satellites that are not yet disrupted, such as the \LMC{}, are more likely to have a lower mass and more metal poor stellar halo than the overall population of galaxies of similar mass. Eventually, the destruction of the massive satellite leads to an increase in the stellar halo mass and metallicity.
The solution to the MW's atypical properties is provided by a fourth unusual feature of the \MW: the presence of a satellite with a mass as large as the \LMC's.}

Using a semi-analytic orbital evolution model that includes the MW,
the LMC and Andromeda, we established that the LMC will likely merge
with our galaxy in $2.4^{+1.2}_{-0.8}~\rm{Gyrs}$ (68\% confidence
level), where the confidence interval has been calculated using a
large number of Monte Carlo realizations that account for
uncertainties in the LMC proper motions and in the dark matter halo
masses of the LMC, MW and Andromeda. More than $93\%$ of the Monte
Carlo realizations end up with a \MWLMC{} merger, and, furthermore,
the merger is insensitive to the presence of Andromeda since most
realizations predict a merger well before our massive neighbour comes
within a distance of $300\kpc$ from the Galaxy.

The \MWLMC{} merger is an inevitable consequence of the large dark
matter mass that the LMC appears to have. Even though the LMC is
currently heading away from the \MW{}, dynamical friction acting on
such a heavy galaxy will cause its orbit rapidly to lose energy and,
approximately a billion years from now, to turn around and head
towards the centre where it is destined to merge in another 1.5
billion years or so.  The high mass of the LMC halo inferred from
dynamical considerations \citep{Penarrubia2016} is consistent
with the fact that in the \Eagle{} simulations, LMC-mass satellites
that themselves have a satellite as massive as the SMC typically have a halo
mass, $M_{200}=3.0^{+0.7}_{-0.8}\times 10^{11}\Msun$, at infall.

A massive LMC alters the position and velocity of the MW barycentre
which, in turn, affects the eventual encounter between the MW and
Andromeda,
\changed{as anticipated by \citet[][]{Gomez2015} and calculated using a simplified model by \citet{Penarrubia2016}.}
Our model shows that the Andromeda collision will be less
head-on than previously thought, and that Andromeda's tangential
velocity with respect to the \MWLMC{} barycentre is higher than
previously estimated, which is in better agreement with
cosmological expectations \citep{Fattahi2016}.
We find that the first
encounter between the MW and Andromeda will take place in
$5.3_{-0.8}^{+0.5}\Gyrs$ (68\% confidence level), which is at least
$1.5\Gyrs$ later than previous estimates.

To discover the likely outcome of the \MWLMC{} merger, we identified
analogue systems in the \Eagle{} simulation and followed their
evolution through the merger process. We selected analogues by
matching the black hole, gas and halo masses of the \MW{} and the
stellar mass of the LMC. The merger of LMC
analogues leads to large growth in the black hole mass of the MW
analogues, with a clear enhancement compared to a merger-free control
sample. Most of the mass of the merging satellite is deposited in the
stellar halo between 10 and $100\kpc$ from the central galaxy. The
merger also imparts gravitational kicks to a significant number of
stars in the central galaxy which join the stellar halo. The
metallicity of the halo is greatly increased.

Grafting the results of the \Eagle{} \MWLMC{} analogues to the real MW
we find that following the merger of the LMC the Galactic black hole
mass will increase by a factor of between 1.5 and 8 (median value
2.5). The merger will not destroy the disc, and the Galactic bulge
will hardly change. This is exactly the trend needed to bring our
Galactic black hole onto the mean black hole--spheroid mass relation.
Debris from the LMC merger will overwhelm the stellar halo, whose mass
will increase by a factor of between 3 and 6 (median value 5). This
will promote the MW from the galaxy with the lowest stellar halo mass
to an average galaxy. The metallicity of the newly formed stellar halo
will effectively be that of the LMC, which is on the high side (but
within the scatter) of the observed stellar halo metallicities in
galaxies similar to the MW. The collision with the \LMC{} will have
restored our Galaxy to normality.

The growth of the supermassive black hole following the future
\MWLMC{} merger will trigger AGN activity and possibly generate jets
which, in turn, can produce powerful $\gamma{-}\textrm{ray}$ emission
\citep[e.g.][]{Padovani2017}. If energetic enough,
$\gamma{-}\textrm{rays}$ impinging on Earth can cause mass extinctions
by destroying the planet's ozone layer \citep{Thomas2005}. However,
the Galactic AGN will not be powerful enough to deplete the Earth's
ozone layer and is very unlikely to pose a serious danger to 
terrestrial life. The \MWLMC{} merger will gravitationally eject
central disc stars into the halo. Is the Sun a potential victim?
Thankfully, this seems unlikely, as only a few percent of the stars at
the position of the Sun in our \MWLMC{} analogues are kicked out into
the halo.

\section*{Acknowledgements} 
\changed{We thank the anonymous referee for their insightful comments.}
We also thank Andrew Cooper, David Rosario and Joop Schaye for very helpful
discussions. MC and CSF were supported by the Science and Technology
Facilities Council (STFC) [grant number ST/I00162X/1, ST/P000541/1];
CSF, was supported, in addition by an ERC Advanced Investigator grant,
DMIDAS [GA 786910].  AD is supported by a Royal Society University
Research Fellowship.  SM was supported by STFC
[ST/F001166/1,ST/L00075X/1] and by Academy of Finland, grant number: 314238.  
This work used the DiRAC Data Centric
system at Durham University, operated by the Institute for
Computational Cosmology on behalf of the STFC DiRAC HPC Facility
(www.dirac.ac.uk). This equipment was funded by BIS National
E-infrastructure capital grant ST/K00042X/1, STFC capital grants
ST/H008519/1 and ST/K00087X/1, STFC DiRAC Operations grant
ST/K003267/1 and Durham University. DiRAC is part of the National
E-Infrastructure.

\bibliographystyle{mnras}
\bibliography{ref_MW_LMC_merger}

\bsp	
\label{lastpage}

\end{document}